\documentclass[aps,prapplied,twocolumn]{revtex4}

\usepackage{bm}
\usepackage{graphicx}
\usepackage{amsmath}

\newcommand{\Ohm}{\Omega}
\newcommand{\Cth}{C_\mathrm{th}}
\newcommand{\Gth}{G_\mathrm{th}}
\newcommand{\NET}{\mathrm{NET}}
\newcommand{\Tel}{T_\mathrm{el}}
\newcommand{\Tbath}{T_\mathrm{bath}}

\newcommand{\Tnoise}{T_\mathrm{noise}}
\newcommand{\Pin}{P_\mathrm{in}}
\newcommand{\Pdiss}{P_\mathrm{diss}}
\newcommand{\kB}{k_\mathrm{B}}
\newcommand{\ie}{\emph{i.e.}}
\newcommand{\etal}{\emph{et al.}}
\newcommand{\Qdot}{\dot{Q}}

\begin{document}

\title{Dispersive thermometry with a Josephson junction coupled to a resonator}

\author{O.-P. Saira$^1$, M. Zgirski$^2$, K.L. Viisanen$^1$, D.S. Golubev$^1$, and J.P. Pekola$^1$}
\address{$^1$Low Temperature Laboratory, Department of Applied Physics, Aalto University School of Science, P.O. Box 13500, 00076 AALTO, Finland \\
$^2$Institute of Physics, Polish Academy of Sciences, al. Lotnik\'ow 32/46, PL 02-668 Warszawa, Poland}

\begin{abstract}
We have embedded a small Josephson junction in a microwave resonator that allows simultaneous dc biasing and dispersive readout. Thermal fluctuations drive the junction into phase diffusion and induce a temperature-dependent shift in the resonance frequency. By sensing the thermal noise of a remote resistor in this manner, we demonstrate primary thermometry in the range from 300~mK to below 100~mK, and high-bandwidth (7.5~MHz) operation with a noise-equivalent temperature of better than 10~$\mathrm{\mu K/\sqrt{Hz}}$. At a finite bias voltage close to a Fiske resonance, amplification of the microwave probe signal is observed. We develop an accurate theoretical model of our device based on the theory of dynamical Coulomb blockade.
\end{abstract}

\maketitle

\section{Introduction}

An apparently simple coupled system of an ultrasmall Josephson junction and a transmission line resonator
exhibits very rich physics which currently attracts a lot of attention.
It has been known for a long time that the current voltage characteristics of such a junction is described by well-developed
$P(E)$ theory \cite{ANO,IN,IG}, which emphasizes the effect of the electro-magnetic environment on 
the fluctuations of the Josephson phase. This theory has been experimentally verified, see e.g. in Refs. \onlinecite{Holst,Saclay1,Pashkin}.
In more recent experiments, the "bright side" effect, \ie, emission of radiation by the junction into the transmission line, has been detected \cite{CB},
and the signs of lasing by a single Cooper pair box into a resonator have been observed \cite{Rimberg}.
The experiments have stimulated a series of theory papers that describe, for example,
non-linear quantum dynamics of the system~\cite{Gramich,Kubala},
emission of entangled photons in two separate resonators~\cite{Armour}, 
antibunching of the emitted photons~\cite{Lep1}, and the full quantum theory of emitted radiation~\cite{Lep2}.

In this Letter, we demonstrate that depending on bias condition an ultrasmall Josephson junction can operate
either as a sensitive noise detector or as a source of photons. We weakly couple the resonator to
the outer transmission line and monitor its resonance frequency and the quality
factor via microwave reflection measurements. We show that at zero bias the
shift of the resonance frequency is inversely proportional to temperature and the junction
operates as an ultra-sensitive thermometer and noise detector. In this regime
we effectively realize a power-to-frequency transducer.   
Applying bias voltage $V$ to the junction we detect amplification of the probe
signal close to the resonance condition, where the Josephson frequency $\omega_J=2eV/\hbar$  matches
the fundamental frequency of the resonator $\omega_r$. In this case the junction 
may operate as an amplifier or as a source of radiation.
We also develop a high-frequency generalization of the $P(E)$-theory and show that 
it describes the experiment fairly well in the whole range of bias voltages studied. Our results hilight the unique properties of an ultrasmall Josephson junction and outline future applications as a thermometer or as a general-purpose radiation and noise detector. In particular, we estimate that a thermal photodetector based on this method of temperature sensing could reach photon-resolving energy resolution in the microwave domain.

\begin{figure}[t!]
\includegraphics[width=\columnwidth]{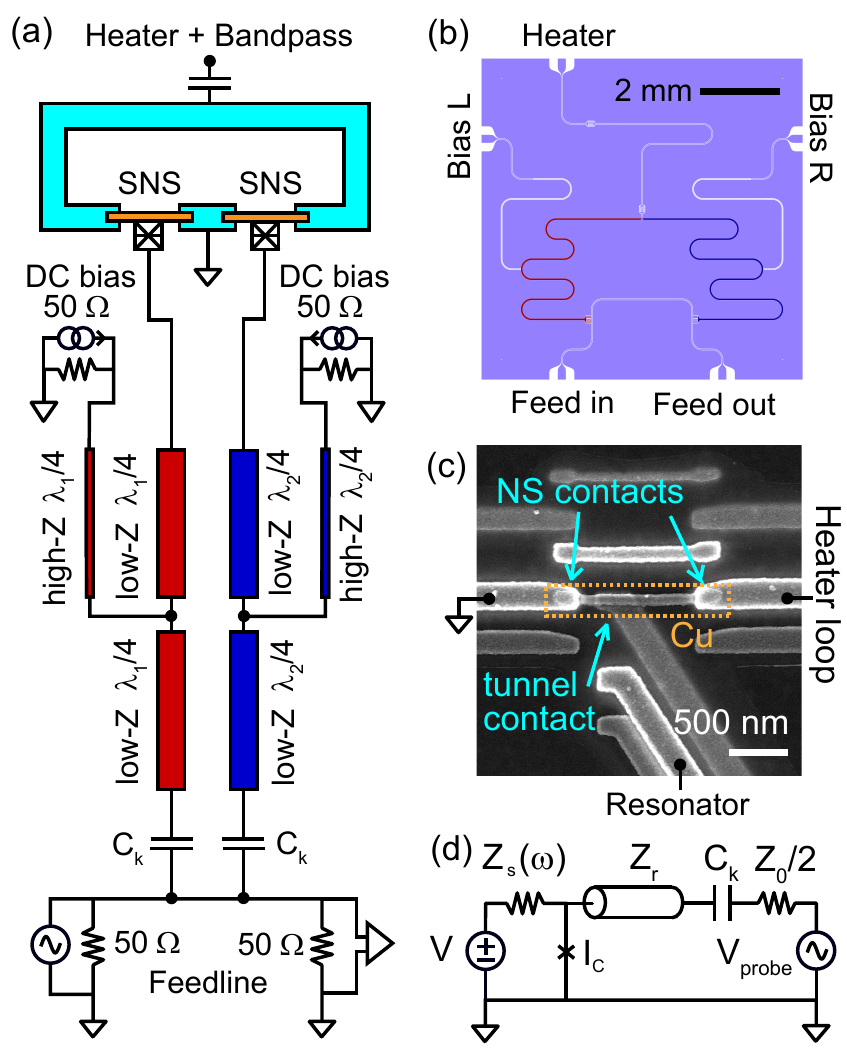}
\caption{The studied system. (a) Electrical schematic of the device and essential external components. The left and right halves of the device are functionally equivalent. Coplanar waveguides are used to define microwave resonance modes (at 4.688~GHz and 5.668~GHz for blue and red elements in the illustration, respectively). Combination of low and high characteristic impedance (30~$\Ohm$ and 125~$\Ohm$, respectively) sections of approximately $\lambda/4$ length allow DC biasing while increasing only slighly the losses of the microwave resonance. The resonators couple capacitively to a common feedline for frequency-multiplexed readout. The resonators terminate at small tunnel junctions between an Al electrode and a proximitized Al/Cu/Al (SNS) wire. The two SNS wires are part of the same superconducting loop~\cite{Foot1} that couples electrical fluctuations from one wire to the other. Lastly, a microwave line is capacitively connected to a section of the loop and can be used to heat the SNS wires. Detailed wiring shown in Supplement~\cite{Sup}. (b) Physical layout of device chip. The CPWs are fabricated from etched Nb on Si substrate. Bandpass filter for the heating line is implemented as an in-line half-wave resonator. (c) Scanning electron micrograph of one of the tunnel junctions fabricated with three-angle shadow evaporation. The disconnected copies of the mask pattern are a byproduct of the fabrication method, and do not affect the device characteristics. (d) Reduced circuit model used in theory. The dc shunt impedance $Z_s(\omega = 0)$ will be denoted by $R_s$.}
\end{figure}

\section{Overview of the experiment}

\subsection{Sample design}

We have fabricated a small Josephson junction and coupled it to a coplanar waveguide (CPW) resonator that allows simultaneous dc biasing and microwave probing [Fig.~1(a),(b)]. The chip layout and the design philosophy of microwave elements mirror those of superconducting quantum processors~\cite{cQED}. In the absence of Josephson dynamics, the fundamental $\lambda/2$ resonance mode is characterized by the resonance frequency $f_r^{(0)}=5.6681$~GHz, the internal and coupling quality factors $Q_i^{(0)}=3400$ and $Q_c = 760$, respectively, and impedance $Z_{lc} = \frac{2}{\pi} Z_r$, where $Z_r = 30~\Omega$ is the characteristic impedance of the waveguide. The Josephson element was realized as a planar tunnel junction between an aluminum electrode and a 1~$\mu$m long proximitized Al/Cu/Al SNS wire [Fig.~1(c)]. A separate heater line allows local Joule heating of the wire to aid in characterization. Earlier experiments on resonator-coupled tunnel junction structures have employed thermometry based on quasiparticle transport~\cite{NIS}, and an initial observation of  supercurrent thermometry was reported in Ref.~\onlinecite{Viisanen}. Details of device fabrication, measurement setup, and microwave readout are presented in the supplement~\cite{Sup}.

The sample chip contains another similar device with the resonator 1~GHz lower in frequency. The two device structures can be independently dc biased and read out by frequency multiplexing, and they showed similar behavior in the experiments. Here, we mainly discuss the higher-frequency device whose readout resonance had smaller intrinsic loss.
 
\subsection{Theory}
 
The junction dynamics is described by the equation
\begin{eqnarray}
\int_{-\infty}^t dt' Y(t-t')\frac{\hbar\dot\varphi}{2e}+I_J=\tilde I_{\rm probe}(t),
\label{RSJ}
\end{eqnarray}
where $Y(t)=\int d\omega\, e^{-i\omega t}/2\pi Z(\omega)$ is the Fourier transformed admittance of
the electromagnetic environment surrounding the junction and including junction capacitance [see Fig.~1(d)], $Z(\omega)$ is the impedance of the environment,
$I_J$ is the Josephson current and $\tilde I_{\rm probe}(t)$ is the current induced by the probe signal. If the junction critical current, $I_c$, is high,
one should put $I_J=I_c\sin\varphi$ in Eq.~(\ref{RSJ}). However, here we consider the limit $I_c\lesssim 2e{\kB}T/\hbar$ in which case, applying the theory developed in Refs.~\onlinecite{Schoen}, we find (see the Supplement~\cite{Sup} for details)
\begin{eqnarray}
I_J&=&\frac{I_c^2}{2e}\int_{-\infty}^t dt' e^{-F(t-t')}\sin[K(t-t')]
\nonumber\\ &&\times\,
\sin\left[\varphi(t)-\varphi(t')+ \omega_J(t-t') \right].
\label{IJ}
\end{eqnarray} 
Here, $\varphi(t)$ is the high frequency component of the Josephson phase induced
by the combined effect of the probe signal and the resonator. 
The functions $F(t)$ and $K(t)$ characterize the environment and are defined as follows
\begin{eqnarray}
F(t) &=& \frac{4e^2}{\hbar^2}\int \frac{d\omega}{2\pi}\,S_V(\omega)\frac{1-\cos\omega t}{\omega^2},
\nonumber\\
K(t) &=& \frac{2e^2}{\hbar}\int\frac{d\omega}{2\pi}\frac{Z(\omega)e^{-i\omega t}}{-i\omega + \epsilon}. \label{eq:FK}
\end{eqnarray}
In these expressions, $S_V(\omega)$ is the spectral density of voltage fluctuations across the junction
and $\epsilon$ is an infinitely small positive constant. In equilibrium, the fluctuation-dissipation theorem is valid and
one finds $S_V(\omega)=\,{\rm Re}\,[Z(\omega)]\hbar\omega\coth({\hbar\omega}/{2{\kB}T})$. In our experiment, the impedance $Z(\omega)$ is dominated by transmission line resonator [see Figs.~1(a) and (d)], and can be formally written as
\begin{eqnarray}
Z(\omega)=i\frac{Z_r}{\pi}\sum_{n=-\infty}^{\infty} \frac{\omega_r^{(0)}}{\omega- n\omega_r^{(0)}+i\gamma_n^{(0)}},
\label{Z}
\end{eqnarray}
where $\omega_r^{(0)}$ is the angular frequency of the fundamental ($\lambda/2$) resonance, and $\gamma_n^{(0)}$ is the damping rate of the $n$th harmonic mode. One has $\gamma_n^{(0)}=\gamma_{i,n}^{(0)}+\gamma_{c,n}^{(0)}$,  where $\gamma_{i,n}^{(0)}=({\omega_r^{(0)} Z_r}/{\pi})\,{\rm Re}\left[{1}/{Z_s(n\omega_r^{(0)})}\right]$ is the internal damping  and $\gamma_{c,n}^{(0)}=n^2 (\omega_r^{(0)})^3C_k^2Z_rZ_0/2\pi$ originates from coupling to the outer transmission line. In the experiment, contributions from up to the second harmonic ($n=2$) can be observed.

\subsection{Linearized treatment}

Linearizing the problem in $\varphi(t)$, we introduce the impedance of the junction
\begin{multline}
Z_J^{-1}(\omega) = i({I_c^2}/{2\hbar\omega}) \times \\
[{\cal P}(\omega_J)+{\cal P}(-\omega_J) - {\cal P}(\omega+\omega_J)-{\cal P}(\omega-\omega_J)],
\end{multline}
where the function 
\begin{eqnarray}
{\cal P}(\omega)=\int_0^\infty dt\,e^{i\omega t}e^{-F(t)}\sin[K(t)]
\label{P00}
\end{eqnarray} 
characterizes the high frequency response of the electromagnetic environment and generalizes the familiar $P(E)$ function.  The latter describes only the DC properties of the junction, \ie, its $I$-$V$ curve. The two functions are related as  ${\rm Im}\,[{\cal P}(\omega)]=\pi\hbar[P(\hbar\omega)-P(-\hbar\omega)]/2$. Taking the limit $Z_r/R_q \ll 1$ and making use of the small-$I_c$ assumption, we find the modified resonance frequency $f_r$ and the internal damping rate $\gamma_{i,1}$ of the fundamental
resonance ($n=1$) as
\begin{eqnarray}
f_r & = & f_r^{(0)} + ({f_r^{(0)}Z_r}/{\pi})\,{\rm Im}\left[Z_J^{-1}(\omega_r^{(0)})\right],\\
\gamma_{i,1} & =&  \gamma_{{i,1}}^{(0)} - (\omega_r^{(0)} Z_r/\pi)\,{\rm Re}\left[Z_J^{-1}(\omega_r^{(0)})\right].
\end{eqnarray}

\section{Zero-bias operation}

\begin{figure}[t!]
\includegraphics[width=\columnwidth]{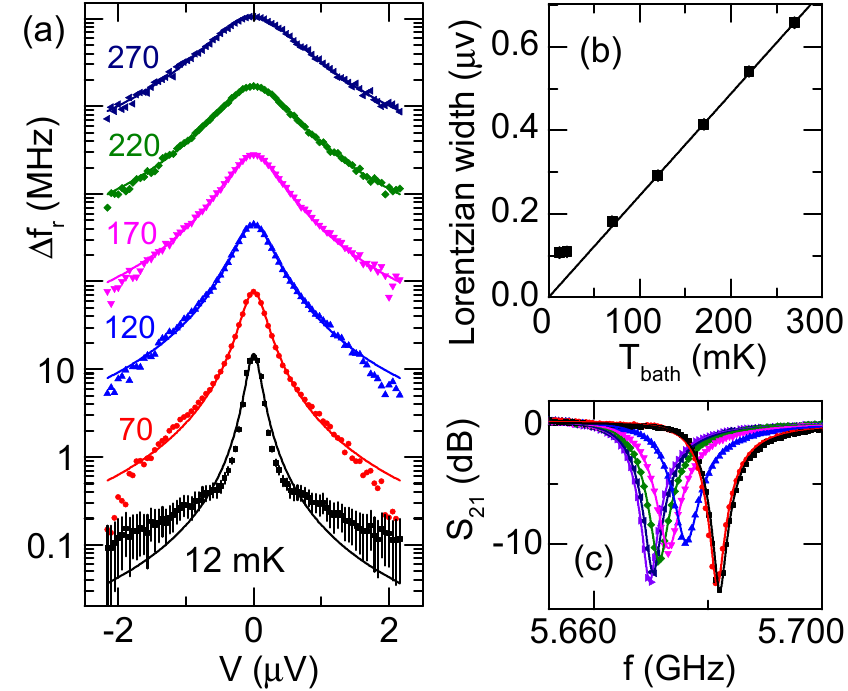}
\caption{Primary thermometry in phase diffusion regime of the Josephson junction. (a) Measured change in the resonance frequency as a function of bias voltage at different bath temperatures relative to a baseline ($\approx$ 5.668~GHz) established from Lorentzian fits (lines). Above 12 mK, data has been vertically offset by one order of magnitude per temperature point for clarity. (b) The width of the zero-bias feature as a function of bath temperature and a linear fit with zero intercept. (c) Resonance lines at $V=0$ at different bath temperatures. Marker symbols (data) and line color (fits) indicate temperature as in panel (a).}
\end{figure}

At low bias voltages and in the limit of small-signal microwave probing, the voltage dependence of the resonance frequency reduces to a simple Lorentzian form
\begin{equation}
f_r = f_r^{(0)} + \delta + \frac{\Delta f_T}{1+{V^2}/{V_T^2}},\; V_T = \frac{4\pi R_s {\kB}T}{eR_q}
\label{Lorentzian}
\end{equation}
where $R_s= Z_s(0)$ denotes effective low frequency shunt resistance, and the other parameters are
\begin{eqnarray}
\Delta f_T = \frac{I_c^2Z_r}{4\pi^2 {\kB}T} \left(2\sinh\frac{\pi {\kB}T}{\hbar\gamma_0}\right)^\alpha,\;\delta = -\frac{\Delta f_T}{1+\hbar\gamma_0/2eV_T},\nonumber
\end{eqnarray}
$\alpha=8 e^2 R_s/2\pi\hbar$, and $\gamma_0=Z_r\omega_r^{(0)}/\pi R_s$. To arrive at Eq.~(\ref{Lorentzian}), we have assumed $\hbar \gamma_0 \gg k_B T$, and the limit of classical phase fluctuations $R_s(0),Z_r,Z_0\ll R_q = h/e^2=25.8$~k$\Omega$. Both conditions are satisfied in our experiment.

In the experiment, the resonance line displays a clear temperature [Fig.~2(c)] and bias dependence. In Fig.~2(a) we analyze the experimental low-bias part of frequency-voltage dependence, which indeed has the Lorentzian form. The width of the Lorentzian is proportional to temperature at $T\gtrsim 70$ mK [Fig.~2(b)]. Comparing the experimental temperature dependence of the width with Eq.~(\ref{Lorentzian}), we determine the low frequency shunt resistance $R_s=57.4~\Omega$. By design, the shunt resistance is given by the external bias resistor (nominally 50~$\Omega$) plus any effective in-line dc resistance including the SNS wire and contacts ($\approx 4~\Omega$ for Cu in normal state). The deviation from the linear dependence at lowest temperatures is due to two independent mechanisms. When the condition $I_c< 2ek_BT/\hbar$ is violated, our model no longer applies and a supercurrent feature with a width close to $I_cR_s$ emerges instead. Besides this, insufficient thermalization can result in saturation of sample temperature. With the linear scaling established earlier, the minimum observed width corresponds to a temperature of 44~mK~\cite{Foot4}.

One can similarly work out the approximate form of the quality factor at low bias voltage. The result reads
\begin{multline}
\frac{1}{Q_i}=\frac{1}{Q_i^{(0)}} + \\\frac{2e I_C^2Z_r^2}{\pi^2 \hbar\omega_r^{(0)} V_T}
\left(2\sinh\frac{\pi k_BT}{\hbar\gamma_0}\right)^{8R_s/R_q}\frac{1}{1+V^2/V_T^2}.
\label{Qi}
\end{multline}
Here we assumed that the thermal linewidth significantly exceeds the damping rate 
of the fundamental resonance~\cite{Sup}. This condition is satisfied in our experiment.

\begin{figure}[t!]
\includegraphics[width=\columnwidth]{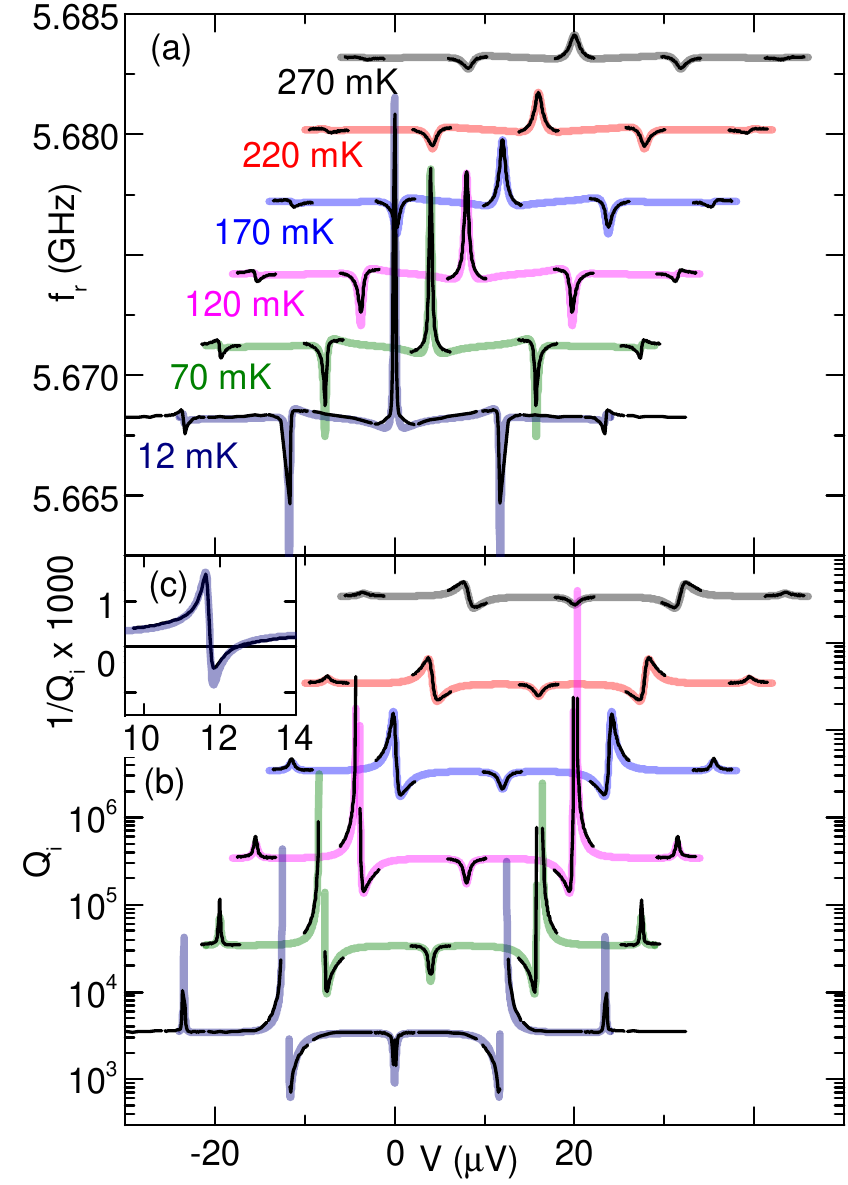}
\caption{Bias dependence of microwave response together with detailed theory. (a), (b) Resonance frequency and quality factor, respectively, versus bias voltage at different temperatures. Data at temperatures other than 12 mK have been offset for clarity. Thin lines are experimental data, thick faded lines are fits with full small-signal theory. We have used $R_s=57.4~\Omega$,  and have chosen the critical currents $I_c=$ 3.25, 3.21, 3.14, 3, 2.7, 2.42, 2.06, nA for the bath temperatures $T=$ 12, 20, 70, 120, 170, 220, and 270 mK respectively. The noise temperature sensed by the junction is set in accordance with the width of the Lorentzian zero bias feature, c.f. Fig.~2(b). The inset (c) highlights a region of negative $Q_i$ observed at base temperature as predicted by theory.}
\end{figure}

\section{Finite-voltage resonances}

In Figs.~3(a), (b) we show, respectively, the resonance frequency, $f$, and the internal quality factor, $Q_i$,
as a function of the bias voltage applied to the junction. The experimental data are fitted with the temperature-dependent critical current $I_c$ as the only free parameter. (Refer to Supplement for comparison of $I_c$ values determined with different methods, and for theory expressions covering full bias range~\cite{Sup}). It is interesting that the internal quality factor $Q_i$ becomes negative at bias voltages close to $\hbar\omega_r^{(0)}/2e$
and at a sufficiently low temperature [Fig.~3(c)]. In this regime the junction pumps energy into the resonator 
and amplifies the probe signal. Previously emission from the junction has been detected under similar conditions \cite{CB}.
The theory predicts that the internal damping becomes negative at $T<T^*=I_c^2Z_rQ_i^{(0)}/4\pi {\kB}\omega_r^{(0)}$ and for
bias voltages in the range 
\begin{eqnarray}
\left|V-\frac{\hbar\omega_r^{(0)}}{2e}-\frac{2e R_s\kB T^*}{\hbar}\right|<\frac{2e R_s\kB\sqrt{T^{*2}-T^2}}{\hbar}.
\end{eqnarray}
Taking $I_c\approx 3$ nA we estimate the threshold temperature to be $T^*\approx 150$ mK. Experimentally, the threshold temperature lies between 120 and 170~mK based on data shown in Fig.~3(b). Although the condition $Q_i < 0$ indicates the generation of microwave power by the junction, 
it does not imply $|S_{21}| > 1$ in the two-port feedline configuration employed in our experiment. For that, a stricter condition $1/Q_i + 1/2Q_c^{(0)} < 0$ needs to be met, which occurs theoretically at $T < T^*/(1+Q_i^{(0)}/2Q_c^{(0)})\approx 45$~mK and was not realized in the experiment. 
In earlier work~\cite{Pasi}, a one-port device based on this principle has been operated as a reflection amplifier at 2.8~GHz.

\section{Non-linear operation}

\subsection{High-power readout}
\label{sec:hipower}

\begin{figure}[t!]
\includegraphics[width=\columnwidth]{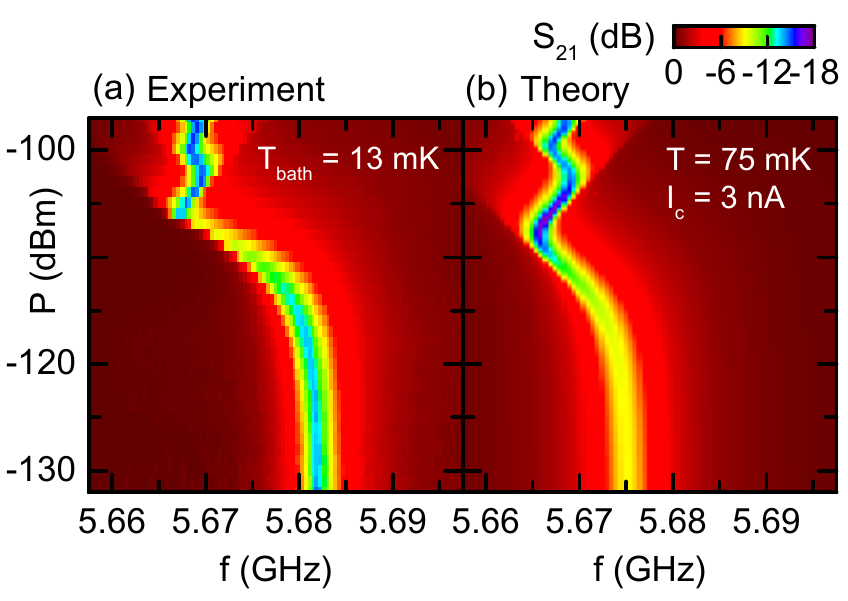}
\caption{Effect of large probe power on resonance line at zero bias. The experimental data (left) was obtained at the base temperature of the cryostat (13 mK). The theory data was calculated using a non-linear model that describes the effect of large phase oscillations on effective junction dynamics. In the simulation, we used an artifically elevated constant temperature (75 mK) for the electromagnetic environment to suppress hysteresis in the model.}
\end{figure}

Here, we relax the assumption $|\varphi| \ll 1$ to describe the response to strong microwave probing. High-power probing is relevant for optimizing the noise-equivalent temperature, although overheating of the sample can impose a stricter limit to probing power than the non-linearity of Josephson dynamics. In a two-port feedline configuration employed in the experiment, the amplitude of high-frequency phase modulation $\phi_1$ is related to the incident probe power $P_\mathrm{in}$ at probe frequency $f_p$ as
\begin{equation}
\phi_1 = \frac{4 Q}{e R_q f_r} \sqrt{\frac{2 Z_r P_\mathrm{in}}{\pi Q_c \left(1 + 4 Q^2 \frac{(f_p - f_r)^2}{f_r^2}\right)}},\label{phi1}
\end{equation}
where $Q^{-1} = Q_c^{-1} + Q_i^{-1}$. Denoting by $\tilde f$ and $\tilde Q_i$ the power-dependent expressions for the resonance frequency and internal quality factor, respectively, we find the relations
\begin{eqnarray}
\tilde f_r - f_r^{(0)} & = & \left[J_0^2(\phi_1) -J_1^2(\phi_1)\right] \left(f_r - f_r^{(0)}\right)\label{frtilde}\\
\frac{1}{\tilde Q_i} - \frac{1}{Q_i^{(0)}} &= & \left[J_0^2(\phi_1) -J_2^2(\phi_1)\right] \left(\frac{1}{Q_i} - \frac{1}{Q_i^{(0)}}\right),\label{Qitilde}
\end{eqnarray}
where the $J_n$ are Bessel functions of the first kind, and the small-signal $f_r$ and $Q_i$ are evaluated according to Eqs.~(4) and (5), respectively. An experimental power sweep performed at zero bias and at the base temperature of the cryostat (13~mK) [Fig.~4(a)] indeed reveals Bessel-type oscillations of the resonance frequency. Solution of the circuit model with $\phi_1$-dependent $\tilde{f}_r$ and $\tilde{Q}_i$ reproduces the data well [Fig.~4(b)] including  fine structure that appears with off-resonant probing at large power.

The non-linearity of the model can result in multi-valued solutions for certain combinations of low temperature, large $I_c$, and large probing power. We did not observe hysteretic or bistable behavior in the experiment. Physically, it is likely that large probing power locally heats up parts of the sample or the surrounding circuitry, raising the effective temperature. It is in principle possible to include a thermal balance in the model and solve it in a self-consistent manner. Here, we explain the high-power response by using a constant elevated temperature (75~mK) throughout the simulation. Good agreement with the constant-temperature simulation shows that the present design is not severly overheated even at $-100$~dBm incident probing power.

\subsection{Local heating}

In an indealized desription of our device, the Josephson element does not have an internal temperature of its own. Instead, the observed temperature dependence stems from fluctuations of the electromagnetic environment. Localized Joule heating or electronic cooling of the SNS wire will generally drive the system to a quasi-equilibrium state with independent electron and environment temperatures~\cite{Giazotto}. We demonstrate sensitivity to the local electron temperature by modulating the wire temperature with either CW microwave heating, or by voltage biasing the other tunnel junction that was otherwise unused in the experiment~(data shown in the Supplement~\cite{Sup}). The data is consistent with a model where the cryostat sets the temperature of electromagnetic environment by thermalizing the cold bias resistor, and the wire temperature is probed through its effect on the $I_c$ of the junction. Here, the temperature dependence follows from that of proximity superconductivity in diffusive metallic weak links~\cite{Dubos}. Optimized detectors based on this  mode of operation have been explored in detail in earlier works by Govenius \etal~\cite{Govenius}.

\section{Sensitivity and noise}

\begin{figure}[t!]
\includegraphics[width=\columnwidth]{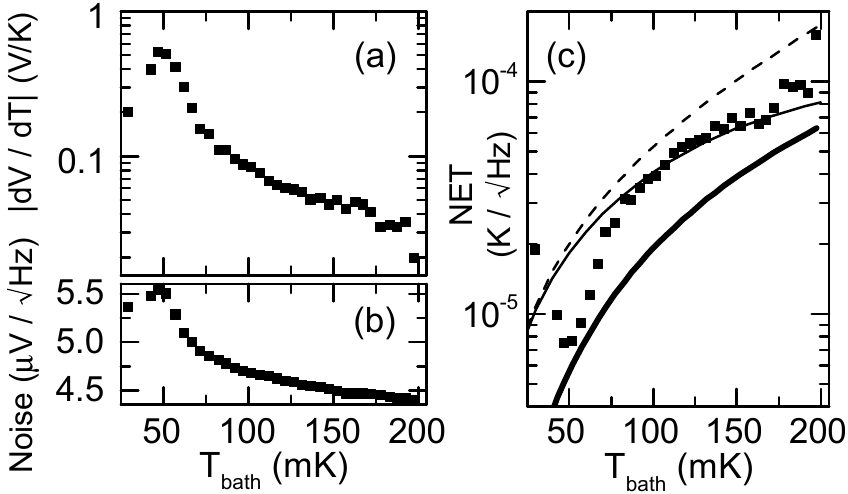}
\caption{Sensitivity and noise of temperature readout with quadrature detection. (a) Voltage responsivity at the output of the readout chain using heterodyne detection with $f_\mathrm{IF}$ = 1 MHz. Magnitude of the complex quantity is shown. (b) Mean spectral density of voltage noise in the frequency band from 10~kHz to 0.3 MHz. (c) NET inferred from (a) and (b) including a $\sqrt{2}$ gain from ideal homodyne detection (experiment, symbols) with superimposed theoretical models: Constant probing power in small-signal regime (dashed line). Same with responsivity enhancement from explicit temperature dependence of $I_c$ (thin solid line). Minimum achievable NET with optimized probing power at each temperature from a non-linear model (thick line).}
\end{figure}

To evaluate the suitability of this thermometer for calorimetric and bolometric experiments~\cite{Foot2}, we characterize the sensitivity of the temperature readout with continuous wave (CW) microwave probing at zero bias with phase-sensitive heterodyne readout. Despite conceptual similarities with noise thermometry employing SQUID readout~\cite{Schwab}, our device indicates temperature directly through a change in the phase of the probe signal instead of relying on power detection with room-temperature electronics. To scan rapidly the parameter space of possible combinations of probing frequency and power, we studied the single-shot detection fidelity of discrete heating pulses (1~$\mu$s duration, 0.8~pW nominal power) using only the thermometer readout. An optimum was found at 5.671~GHz, nominal $-$118~dBm power incident at the sample box. Next, during a bath temperature sweep up to 200~mK, we recorded the CW quadrature voltage amplitudes $V_I$, $V_Q$ and the full noise spectrum of the quadrature readout.
We evaluate numerically the voltage responsivity $\frac{dV}{dT} = \sqrt{\left(\frac{dV_I}{dT}\right)^2 + \left(\frac{dV_Q}{dT}\right)^2}$ corresponding to homodyne detection with optimal phase [Fig.~5(a)]. Similarly, the NET for homodyne detection is $V_{rms} \left(\frac{dV}{dT}\right)^{-1} / \sqrt{2}$, where $V_{rms}$ is the voltage noise level in one quadrature [Fig.~5(b), (c)]. Using the small-signal theory and sample parameters determined earlier, we can reproduce the observed NET values for temperatures higher than 75~mK. We have included the responsivity enhancement from weak temperature dependence of $I_c$ in the model. Comparing the results to a calculation with $\frac{\partial I_c}{\partial T} = 0$, we find that inductance and noise contributions to responsivity are equal at 200~mK, with the inductance modulation losing its significance below 100~mK. For the theoretical NET calculation, we have assumed a total power loss of 11~dB from cabling between generator output and the cold amplifier, and amplifier-limited system noise with $T_\mathrm{noise} = 2$~K (as per preamplifier specifications). These quantities cannot be independently determined within a linearized model. The origin of the temperature dependent component of readout noise that follows the shape of the responsitivity curve is unknown. The low frequency resonator was measured simultaneously in an identical manner and the noise level was found to be constant within 0.5\%. We estimate the power dissipated at the sample ($\Pdiss$, including shunt resistors) as $\eta\Pin$, where $\Pin$ is the incident probing power and $\eta = 2 Q_c Q_i/(Q_c + Q_i)^2$, using modeled values for $Q_c$ and $Q_i$. The loss fraction $\eta$ is smaller than 0.5 at all temperatures, resulting in total dissipation less than 0.8~fW~\cite{Foot3}. Finally, using the theory for high-power readout presented in Sec.~\ref{sec:hipower}, we evaluate the lowest achievable NET when overheating of the sample is neglected~[Fig.~5(c), thick line].

\section{Outlook}

Small power dissipation, sub-$\mu$s temporal resolution, and good sensitivity at sub-100 mK temperatures make this type of a themometer a promising candidate for calorimetric experiments~\cite{Foot2}. In a nano-calorimeter implementation~\cite{Viisanen}, the external macroscopic bias resistor would be replaced with a metallic or semiconducting nanowire with similar resistance but minimal volume. In a calorimeter device, it is critical to consider the tradeoff between the thermometer sensitivity and the power dissipation induced by the thermometer readout. One can formalize this tradeoff by writing the noise-equivalent temperature (NET, units $\mathrm{K/\sqrt{Hz}}$) explicitly in terms of $\Pin$. For a dispersive thermometer, the general result
\begin{equation}
\mathrm{NET} = \frac{f_r Q_c}{4Q^2} \left(\frac{df_r}{dT}\right)^{-1} \sqrt{\frac{k_B \Tnoise}{\Pin}}
\end{equation}
follows from linearized circuit theory assuming one-port reflection measurement and readout noise that is described by the system noise temperature $\Tnoise$. As long as the responsivity $\frac{df_r}{dT}$ does not explicitly depend on $Q_c$, the choice $Q_c = Q_i$ is optimal. For a pure reflection measurement, this implies $\Pin$ = $\Pdiss$ at resonance. It is possible to derive simple expressions describing our Josephson thermometer by substituting the linear-response formulas of Eqs.~(\ref{Lorentzian}) and (\ref{Qi}) with $V=0$ and assuming $R_s/R_q \ll 1$. One has
\begin{equation}
\NET = \frac{2 T Z_r}{\pi R_s} \sqrt{\frac{\kB \Tnoise}{\Pin}}\label{NETjj},
\end{equation}
to the first order in $\frac{Z_r}{\pi R_s}$. Working from the above relation, one can estimate the expected energy resolution of a calorimeter under quite general assumptions (see Appendix A for details) about the temperature dependence of the heat capcity ($\Cth \propto T^a$) and the thermal link of the calorimeter platform ($\Gth \propto T^{b-1}$) as 
\begin{equation}
\delta E \approx 2\sqrt{a + b + 1}\ \sqrt{\Cth T\ \kB \Tnoise}\ \frac{Z_r}{\pi R_s}.
\end{equation}
Note that the validity of Eq.~(\ref{Lorentzian}) requires the fraction $\frac{Z_r}{\pi R_s}$ to be larger than $\kB T/\hbar \omega_r$. For a practical example, we consider a small metallic absorber ($\Cth = 300\,\kB$, $a=1$) on a suspended platform with quantized phononic heat conductance ($b = 2$, see Ref.~\onlinecite{Schwab}) at a temperature of 20~mK, microwave probe at $\omega_r = 2\pi \times 5$~GHz, and a readout chain approaching the standand quantum limit $\kB \Tnoise = \hbar \omega_r$, which results in an estimated energy resolution of $9.6~\mathrm{GHz}\times h$.

In conclusion, we have constructed a power-to-frequency transducer based on a small Josephson junction and demonstrated sensitive high-bandwidth thermometry at sub-100 mK temperatures. We have also developed a theoretical model based on strong environmental fluctuations that describes the measurements within its expected range of validity. Good performance and versatility of the approach suggest it can find use in a wide range of experiments requiring sensitive thermometry, calorimetry, or noise detection. Our results also hint at the possibility of further performance gains in designs with large $I_c$ and/or $R_s$, whose analysis, however, requires an improved theoretical model.

\mbox{}
\begin{acknowledgments}
This work was funded through Academy of Finland grants no. 2722195, 284594, and 285300. We acknowledge the availability of the facilities and technical support by Otaniemi research infrastructure for Micro and Nanotechnologies (OtaNano), and VTT technical research center for sputtered Nb films. M.Z. thanks the \mbox{EAgLE} project. K.L.V. acknowledges financial support from Jenny and Antti Wihuri foundation. We thank A. Savin for the dilution refrigerator setup.
\end{acknowledgments}

\appendix

\section{Calorimeter optimization}

We consider a generic calorimeter platform that is described by the model equations
\begin{eqnarray*}
\delta E &= &\NET \sqrt{\Cth \Gth}\\
\Cth & = & A T^a\\
\Gth & = & \frac{\partial \Qdot}{\partial T}\\
\Qdot & = & B (T^b - \Tbath^b)\\
\NET & = & f(\Pdiss, T)
\end{eqnarray*}
subject to steady-state thermal balance
\begin{equation}
\Pdiss - \Qdot = 0,
\end{equation}
where $T$ and $\Tbath$ denote the temperature of the calorimeter and its surrounding thermal bath, respectively, $\delta E$ is the energy resolution, $\Cth$ is the heat capacity of the calorimeter, $\Gth$ is its linearized heat coductance at the operation point, $\dot{Q}$ describes the steady-state heat flow between the calorimeter and its surroundings, and $a$, $b$ are numbers and $A$ and $B$ numerical constants describing the thermal properties of the calorimeter, and $f$ describes the sensitivity of the thermometer as function of the steady-state dissipation $\Pdiss$ and $\Tel$. The choice of the readout power (or, equivalently, $\Pdiss$) influences the steady-state operation tempeature $T$, and, consequently, $\delta E$ through the temperature dependence of $\Gth$ and $\Cth$.

If one furthermore has
\begin{equation}
f = F \Pdiss^{-1/2} T^c,
\end{equation}
with $F$ a numerical constant and $c$ a number, the problem can be solved through the introduction of a Lagrange multiplier. Note that Eq.~(\ref{NETjj}) describing our thermometer is of this form. One finds
\begin{equation}
\Pdiss^* = \left(a+b + 2c -1\right)^{-1} T^* \Gth^*
\end{equation}
and
\begin{equation}
\delta E^* = \left(a+b+2c-1\right)^{1/2} F \ (T^*)^{c-1/2} \sqrt{\Cth^*},\label{eq:Estar}
\end{equation}
where the superscript $*$ denotes quantities calculated at the optimum steady-state $(\Pdiss, T)$ operation point. In practice, one can evaluate  Eq.~(\ref{eq:Estar}) with $\Tbath = T^*$ to approximate the energy resolution, as the optimal probing power does not raise the absorber temperature significantly.


\begin{thebibliography}{99}
\bibitem{Foot1} Weak flux modulation with a global coil was observed. For the higher, more sensitive resonator, the peak-to-peak amplitude was 2~MHz at zero bias. In all subsequent experiments, flux setting for the largest frequency shift was used.
\bibitem{Foot3} Cabling losses on the input side are not included, and would diminish the estimated dissipation proportionally.
\bibitem{Foot2} Here, we refer to an experimental setting where radiation is coupled to a small absorber, and the thermometer is used to monitor the absorber temperature.
\end{thebibliography}
\end{document}


\title{Dispersive thermometry with a Josephson junction coupled to a resonator\\SUPPLEMENTAL MATERIAL}

\author{O.-P. Saira$^1$, M. Zgirski$^2$, K.L. Viisanen$^1$, D.S. Golubev$^1$, and J.P. Pekola$^1$}
\address{$^1$Low Temperature Laboratory, Department of Applied Physics, Aalto University School of Science, P.O. Box 13500, 00076 AALTO, Finland \\
$^2$Institute of Physics, Polish Academy of Sciences, al. Lotnik\'ow 32/46, PL 02-668 Warszawa, Poland}

\maketitle

\section{Theory}

\subsection{Josephson current in an ultrasmall junction}

In this section we briefly outline the derivation of the expression for the Josephson current (Eq. (2) of the main text).
We start from the general expression for the Josephson current of a junction embedded in the electromagnetic
environment with the impedance $Z(\omega)$ in the form of the path integral over the
fluctuating phases defined of the forward, $\varphi_1$, and backward, $\varphi_2$, branches of the Keldysh contour (see e.g. the review \cite{SZ}):
\begin{eqnarray}
I_J &=& \int {\cal D}\varphi_1 {\cal D}\varphi_2\; I_C \sin[\omega_Jt+\varphi_1(t)]
\nonumber\\ &&\times\,
\exp\left\{\frac{i}{\hbar}\int_{t_0}^t dt' 
\left( E_J\cos[\omega_Jt'+\varphi_1(t')] - E_J\cos[\omega_Jt'+\varphi_2(t')] + \frac{\hbar \tilde I_{\rm probe}(t')\varphi^-(t')}{2e} \right)\right\}
\nonumber\\ &&\times\,
\exp\left\{-\frac{i}{\hbar}\int_{t_0}^t dt'\int_{t_0}^{t'} dt'' \frac{\hbar\varphi^-(t')}{2e}Y(t'-t'')\frac{\hbar\dot\varphi^+(t')}{2e}
-\frac{1}{2\hbar^2}\int_{t_0}^t dt'\int_{t_0}^{t} dt'' \frac{\hbar\varphi^-(t')}{2e}G(t'-t'')\frac{\hbar\varphi^-(t')}{2e} \right\}.
\label{Int1}
\end{eqnarray}
Here the initial time $t_0$ tends to $-\infty$, $E_J=\hbar I_C/2e$, $\varphi^+=(\varphi_1+\varphi_2)/2$, $\varphi^-=\varphi_1-\varphi_2$, and
\begin{eqnarray}
Y(t)=\int\frac{d\omega}{2\pi}\,\frac{e^{-i\omega t}}{Z(\omega)},\;\;
G(t)=\int\frac{d\omega}{2\pi}\,\frac{S_V(\omega)}{|Z(\omega)|^2}\,e^{-i\omega t}.
\end{eqnarray}
The boundary conditions for the integral (\ref{Int1}) read $\varphi_1(t)=\varphi_2(t)=\varphi(t)$. The symbolic path integral 
${\cal D}\varphi_1 {\cal D}\varphi_2$ also includes the integral over the initial values of the phases $\varphi_1(t_0),\varphi_2(t_0)$
weighted with the initial density matrix of the junction $\rho_0(\varphi_1(t_0),\varphi_2(t_0))$. 
In the limit $E_J\lesssim k_BT$ we can expand the path integral in powers of $E_J$. The lowest order term equals to zero because
in our system the averaged square of phase fluctuation is infinite, $\langle \delta\varphi^2\rangle\to\infty$.
The second order correction reads
\begin{eqnarray}
I_J &=& i\frac{I_C^2}{2e}\int_{t_0}^t dt'\int {\cal D}\varphi_1 {\cal D}\varphi_2\; \sin[\omega_Jt+\varphi_1(t)]
\big( E_J\cos[\omega_Jt'+\varphi_1(t')] - E_J\cos[\omega_Jt'+\varphi_2(t')] \big)
\exp\left\{ i\int_{t_0}^t dt' \frac{\tilde I_{\rm probe}}{2e}\varphi^-\right\}
\nonumber\\ &&\times\,
\exp\left\{-\frac{i}{\hbar}\int_{t_0}^t dt'\int_{t_0}^{t'} dt'' \frac{\hbar\varphi^-(t')}{2e}Y(t'-t'')\frac{\hbar\dot\varphi^+(t')}{2e}
-\frac{1}{2\hbar^2}\int_{t_0}^t dt'\int_{t_0}^{t} dt'' \frac{\hbar\varphi^-(t')}{2e}G(t'-t'')\frac{\hbar\varphi^-(t')}{2e} \right\}.
\label{Int2}
\end{eqnarray} 
Evaluating this Gaussian path integral we arrive at the formula (2) of the main text.

\subsection{Function ${\cal P}(\omega)$}

In our setup the environment of the junction is characterized by the impedance
\begin{eqnarray}
Z(\omega)=i\frac{Z_r}{\pi}\sum_{n=-\infty}^{\infty} \frac{\omega_r}{\omega- n\omega_r+i\gamma_n}.
\label{Z}
\end{eqnarray}
The damping factor of the $n$th resonance is expressed as $\gamma_n=\gamma_{i,n}+\gamma_{c,n}$,  where $\gamma_{i,n}=({\omega_r Z_r}/{\pi})\,{\rm Re}\left[{1}/{Z_s(n\omega_r)}\right]$ is the internal damping  and $\gamma_{c,n}=n^2 \omega_r^3C_k^2Z_rZ_0/2\pi$ originates from coupling to the outer transmission line. Hence the response function of the environment, $K(t)$, reads
\begin{eqnarray}
K(t)=\frac{2e^2}{\hbar}\int\frac{d\omega}{2\pi}\, \frac{Z(\omega)}{-i\omega+\eta} \,e^{-i\omega t}
= -i\frac{2e^2}{\hbar}\frac{Z_r\omega_r}{\pi}\theta(t)\sum_{n=-\infty}^{\infty} \frac{1-e^{-i(n\omega_r-i\gamma_n)t}}{n\omega_r-i\gamma_n}.
\label{K}
\end{eqnarray}  
The function describing the phase fluctuations, $F(t)$, can be split into two parts: the contribution of
low frequency fluctuations $F_0(t)$ and the contribution of high frequency resonances $F_r(t)$. 
We consider the limit of classical phase fluctuations $Z_s(0),Z_r,Z_0\ll R_q = h/e^2=25.8$ k$\Omega$, and $k_BT\ll \hbar\gamma_0$ relevant for
our experiment. Besides that we assume that the resonances are very narrow and $\gamma_n\ll\omega_r$. 
In this case we find
\begin{eqnarray}
F_0(t) = \frac{8Z_r\omega_r}{\pi R_q\gamma_0}\ln\left[ \frac{\sinh(\pi k_BT_{\rm env}t/\hbar)}{\sinh(\pi k_BT_{\rm env}/\hbar\gamma_0)} \right] 
\approx \Gamma t -  \frac{8R_s}{R_q}\ln\left[ \sinh(\pi k_BT_{\rm env}/\hbar\gamma_0) \right],
\end{eqnarray}
where $\gamma_0=Z_r\omega_r/\pi R_s$,
\begin{eqnarray}
\Gamma=\frac{8\pi R_s}{R_q}\frac{k_BT_{\rm env}}{\hbar}
\label{Gamma}
\end{eqnarray}
is the thermal linewidth induced by low frequency fluctuations in the environment, $R_s= Z_s(0)$ is the effective low frequency shunt resistance, 
$T_{\rm env}=S_V(0)/2k_BR_s$ is the effective temperature of the environment.
The contribution of the resonator may be expressed in the form
\begin{eqnarray}
F_r(t) = \frac{8Z_r}{R_q}\sum_{n=1}^{n_{\max}} \coth\frac{n\hbar\omega_r}{2k_BT_r}\frac{1-\cos (n\omega_rt)}{n},
\end{eqnarray} 
where $n_{\max}$ is the number of sharp resonances  and $T_r$ is the effective temperature of the resonator. Linearizing the problem in $\varphi(t)$, we introduce the impedance of the junction
\begin{eqnarray}
Z_J^{-1}(\omega) &=& i({I_c^2}/{2\hbar\omega})[{\cal P}(\omega_J)+{\cal P}(-\omega_J) - {\cal P}(\omega+\omega_J)-{\cal P}(\omega-\omega_J)],
\end{eqnarray}
where the function 
\begin{eqnarray}
{\cal P}(\omega)=\int_0^\infty dt\,e^{i\omega t}e^{-F(t)}\sin[K(t)]
\label{P00}
\end{eqnarray} 
characterizes the high frequency response of the electromagnetic environment and generalizes the familiar $P(E)$ function. The two functions are related as  ${\rm Im}\,[{\cal P}(\omega)]=\pi\hbar[P(\hbar\omega)-P(-\hbar\omega)]/2$. The latter describes only the DC properties of the junction, \ie, its $I$-$V$ curve. Since in our sample $Z_r\ll R_q$, we find $F_r(t)\ll 1$,  $K(t)\ll 1$, which allows us to approximate ${\cal P}(\omega)$ as
\begin{eqnarray}
{\cal P}(\omega)=\int_0^\infty dt\, e^{i\omega t} e^{-F_0(t)} K(t).
\end{eqnarray}  
Evaluating the time integral we find
\begin{eqnarray}
{\cal P}(\omega) = \frac{2e^2 Z_r}{\pi\hbar}\left(2\sinh\frac{\pi k_BT}{\hbar\gamma_0}\right)^{8R_s/R_q}
\sum_{n=-\infty}^\infty \frac{\omega_r}{n\omega_r-i\gamma_n} 
\left(\frac{1}{\omega+i\Gamma} - \frac{1}{\omega-n\omega_r + i(\Gamma+\gamma_n)}\right).
\label{P}
\end{eqnarray}
We derive the resonance frequencies from the condition
\begin{eqnarray}
Z^{-1}(\omega) + Z_J^{-1}(\omega) = 0.
\label{disp}
\end{eqnarray} 
Considering $I_c^2$ as a small parameter, we solve Eq.~(\ref{disp}) approximately
and find the modified resonance frequency $f_r$ and the internal damping rate $\gamma_{i,1}$ of the fundamental
resonance ($n=1$) as
\begin{eqnarray}
f_r &=& f_r^{(0)} + ({f_r^{(0)}Z_r}/{\pi})\,{\rm Im}\left[Z_J^{-1}(\omega_r^{(0)})\right],
\label{f}\\
\gamma_{i,1} &=& \gamma_{{i,1}}^{(0)} - (\omega_r^{(0)} Z_r/\pi)\,{\rm Re}\left[Z_J^{-1}(\omega_r^{(0)})\right].
\label{gamma}
\end{eqnarray}

\section{Experiment details}

\subsection{Sample fabrication}

A polished Si substrate was treated with an ammonium fluoride etching mixture (AF 90-10) to remove surface oxides and subsequently coated by exposure to HMDS vapour. 200~nm of Nb was sputtered on the substrate. The waveguide design was transferred to AR-P6200 resist using e-beam lithography, and the underlying Nb film was patterned through reactive-ion etching with SF$_6$-based chemistry. Another e-beam lithography round was performed on a stack of PMMA and P(MMA-MAA) copolymer resists. In an e-beam evaporator chamber, the sample was subjected to Ar ion milling to clean the contact areas in the Nb film. In the same vacuum cycle, three metal films were deposited using e-beam evaporation at variable angles: 15~nm Al (oxidized at 1 mbar O$_2$ for 90~s), followed by 20~nm Cu, followed by 50~nm Al. In this manner, tunnel junctions were formed between the first Al film and the Cu film, and transparent contacts between  Nb and the first Al film, and between Cu and the second Al film.

\subsection{Low-temperature setup}

\begin{figure}[t]
\includegraphics[width=160mm]{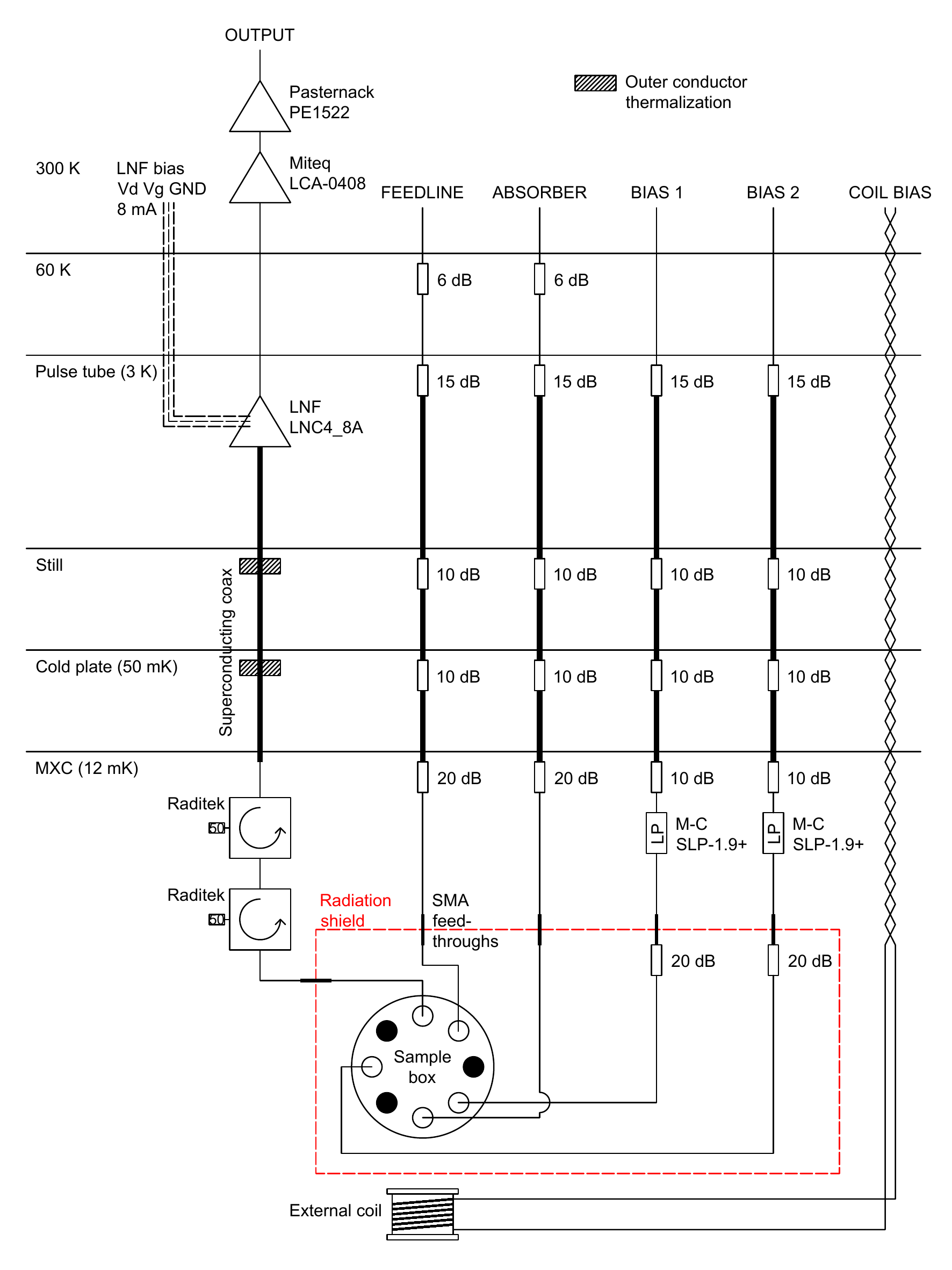}
\caption{Diagram of microwave and twisted pair wiring inside the cryostat relevant to the experiment, and the microwave amplification chain.\label{Fig:Wiring}}
\end{figure}

The sample chip was cooled down in a BlueFors BF-LD250 dilution refrigerator. Electrical signals to the sample (microwave probe and heating signals, and the junction dc biases) were delivered through attenuated 50~$\Ohm$ coaxial lines~[see Fig.~(\ref{Fig:Wiring})]. In particular, the cold resistors that formed the dominant low-frequency electromagnetic environment for the junctions were 20~dB cryogenic attenuators (XMA Corp.) that were located inside a radiation shield at the mixing chamber stage and connected to the sample box with 10~cm long flexible coax lines. External magnetic field was applied with a superconducting coil in the immediate vicinity of the sample box. Temperature of the mixing chamber plate was monitored with a ruthenium oxide resistor calibrated by the cryostat manufacturer, and stabilized with a PI feedback loop.

\subsection{Microwave readout and signal processing}

\begin{figure}[t]
\includegraphics{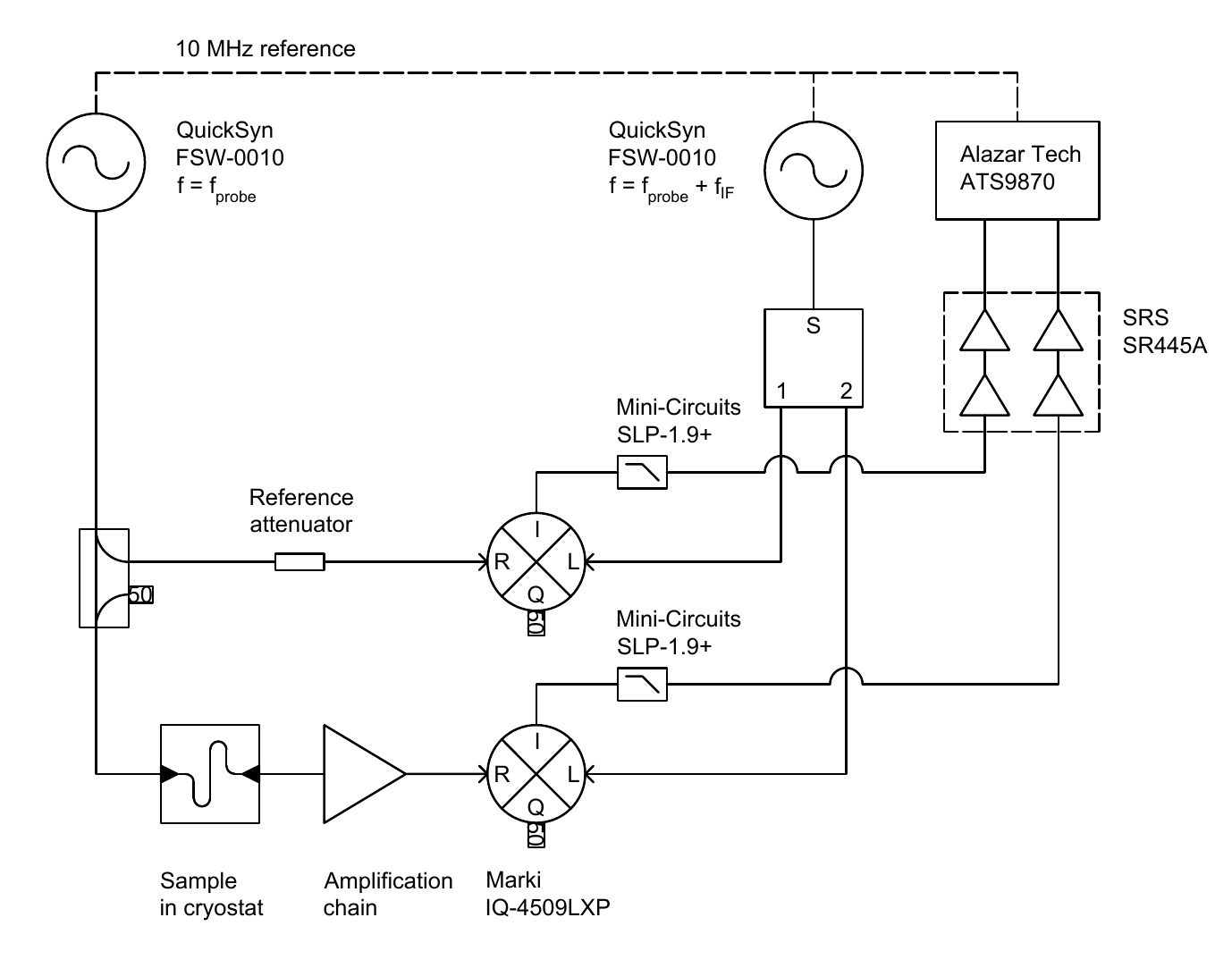}
\caption{Schematic of room-temperature electronics used for quadrature detection.\label{Fig:Detection}}
\end{figure}

Readout was performed with phase-coherent heterodyne detection~[Fig. (\ref{Fig:Detection})]. Two phase-locked microwave generators were used to produce a probe tone at frequency $f_\mathrm{probe}$ and a local oscillator tone at $f_\mathrm{LO} = f_\mathrm{probe} + f_\mathrm{IF}$. The probe tone was split into a reference branch and the measurement branch. Outputs of both branches were down-mixed with copies of the LO tone, low-pass filtered ($f_c = $ 1.9~MHz), and digitized. For steady-state characterization, we chose $f_\mathrm{IF} = 1$~MHz, acquired a long (1 to 100~ms) trace from both digitizer channels simlutaneously, and evaluated complex-valued quadrature phasors $V_\mathrm{meas,ref} = \frac{1}{N}\sum_{k=0}^{N-1} y_\mathrm{meas,ref}[k] \exp \left(-2\pi i k f_\mathrm{IF}/f_{s} \right)$, where $N$ is the number of samples, $f_s$ is the sampling rate, and $y_\mathrm{meas,ref}[k]$ is the $k$th digitizer sample in the acquired from the measurement (reference) channel. The transmission coefficent $S_{21}(f_\mathrm{probe})$ through the measurement branch is then given by $10^{-A/20} V_\mathrm{meas}/V_\mathrm{ref}$, where $A$ is the value of the reference attenuator in dB. We verified that the $S_{21}$ data obtained in this manner agreed with the output of a commercial stand-alone vector network analyzer.

\begin{figure}[t]
\includegraphics[width=172mm]{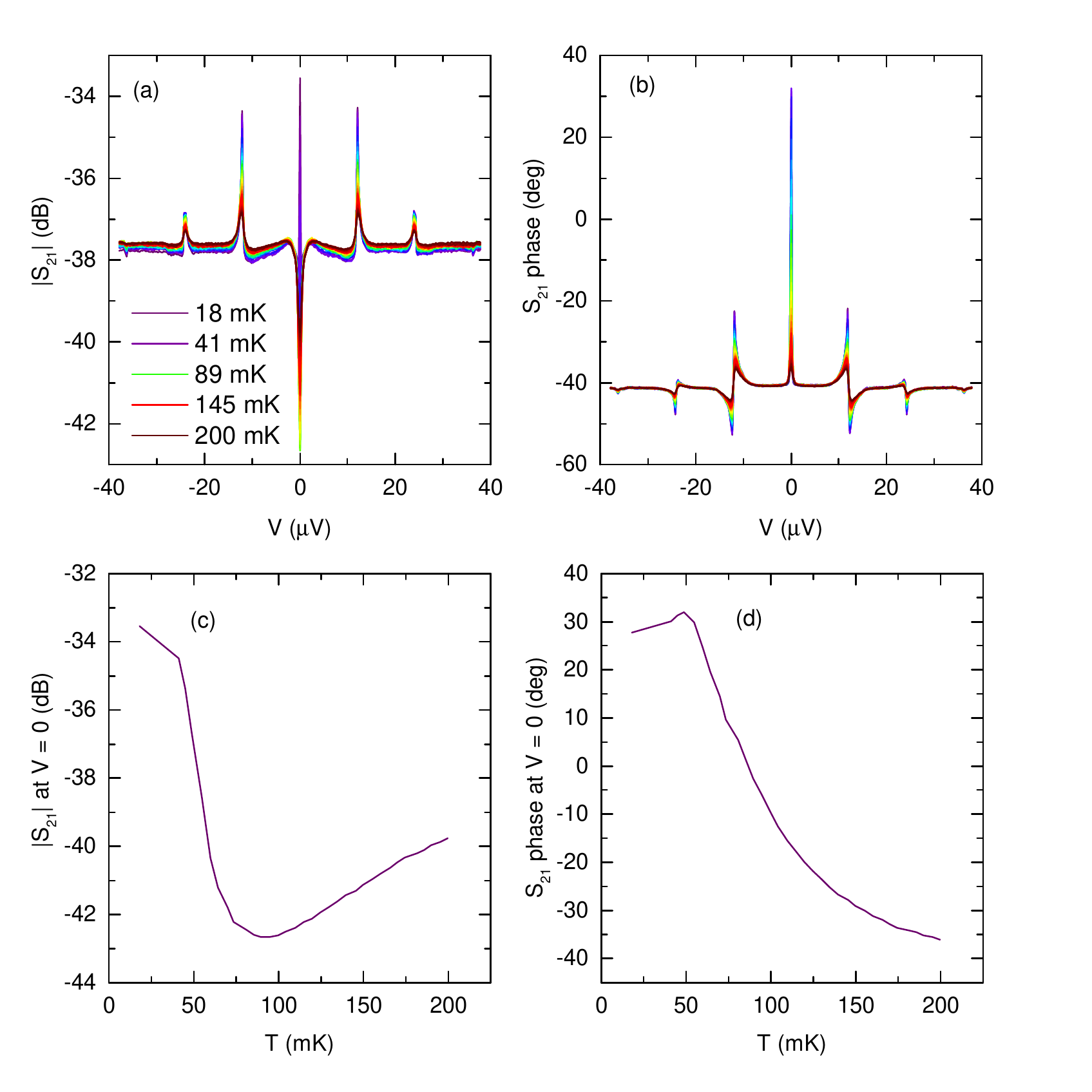}
\caption{Temperature respositivity with constant probing frequency and power. (a), (b) Magnitude and phase of $S_{21}$ as a function of bias voltage at different bath temperatures (34 temperature points in total, see legend for colors). (c), (d) Magnitude and phase of $S_{21}$ at $V=0$ versus bath temperature. \label{Fig:Resp}}
\end{figure}

For estimation of the noise-equivalent temperature, the noise level needs to be calculated in a manner that is consistent with the detection method. To achieve this, we evaluated the discrete fourier transform of digitizer traces in the \emph{same configuration} that was used to determine the $S_{21}$ parameter, and evaluated SNR (signal-to-noise ratio) as the ratio of spectral power in the IF tone at 1~MHz to the average spectral power in the frequency bins from 10~kHz to 0.3~MHz. The quantities plotted in Fig.~5(a) and (b) of the main text are then $\left|V_\mathrm{ref}\right|\times\left|\frac{dS_{21}}{dT_\mathrm{bath}}\right|$ and $\left|V_\mathrm{ref}\right| / \sqrt{f_s\times\mathrm{SNR}}$, respectively, where the derivate with respect to $T_\mathrm{bath}$ is calculated numerically from data obtained during a cryostat temperature sweep. To compensate for possible slow drifts in the total gain and phase of the network, we also recorded the $S_{21}$ parameter in a 50~MHz range around the resonance at each temperature point, and used linear interpolation from the endpoints of this range to calculate a baseline $S_{21}$ at the actual probing frequency. Changes in the estimated baseline $S_{21}$ were compensated before calculation of the derivative. Full voltage dependence of $S_{21}$ obtained during this temperature sweep is shown in Fig.~\ref{Fig:Resp} along with cuts at $V = 0$.

We used homodyne detection ($f_\mathrm{IF} = 0$, no reference branch, both $I$ and $Q$ outputs of the measurement branch mixer are recorded) to rapidly optimize the readout frequency and power by single-shot detection of heating pulses. In this mode, we sent a 1~$\mu$s long heating pulse  ("H1" pulse in the following) with square envelope to the heater line and monitored the response of the thermometer. We calculated integrated voltages $V^H_{I,Q}$ after weighting both quadratures with an exponential decay function ($\tau = 30~\mu\mathrm{s}$) that started $1~\mu$s after the end of the heating pulse. For every second repetition, the heating pulse was disabled to collect an accurate zero reference ("H0" pulse). We collected single-shot histograms of $V^H_{I,Q}$ for at least 1000 repetitions of H0 and H1 cases, calculated the optimal projection angle $\alpha$ that gives the largest separation between them, and finally determined a single threshold voltage $V^*$ which we used to classify the readout traces as "D1" and "D0" (detect 1 and 0, respectively) based on the projected integrated voltages alone. We define the single-shot detection fidelity as $\left[p(D1\,|\,H1) + p(D0\,|\,H0)\right]/2$, \emph{i.\,e.}, the probability of correct classification for an even distribution of H0 and H1 pulses. In a sweep of the frequency of the heating pulse, the detection fidelity showed a clear maximum at 8.38~GHz, yielding the resonance frequency of the on-chip bandpass filter.

\subsection{Fitting of resonance lines}

To extract the frequency and quality factors of the resonance mode, we have fitted the measured resonance lines with a standard one-pole formula~\cite{Geerlings}
\begin{equation}
S_{21}(\omega) = 1 - \frac{\frac{Q}{Q_c} - 2 i Q \frac{\delta \omega}{\omega_r}}{1 + 2 i Q \frac{\omega - \omega_r}{\omega_r}},
\end{equation}
where $\omega$ and $\omega_r$ are the probing and resonance angular frequencies, respectively, $Q^{-1} = Q_c^{-1} + Q_i^{-1}$, and the parameter $\delta \omega$ accounts for a small asymmetry in the resonance line that can be attributed to reflections in the microwave feedline. The least-squares fit was done to complex $S_{21}$ data using additional parameters to describe global amplitude and phase offsets, and a linear phase roll.

\subsection{CPW design and kinetic inductance}

\begin{figure}[t]
\includegraphics[width=150mm]{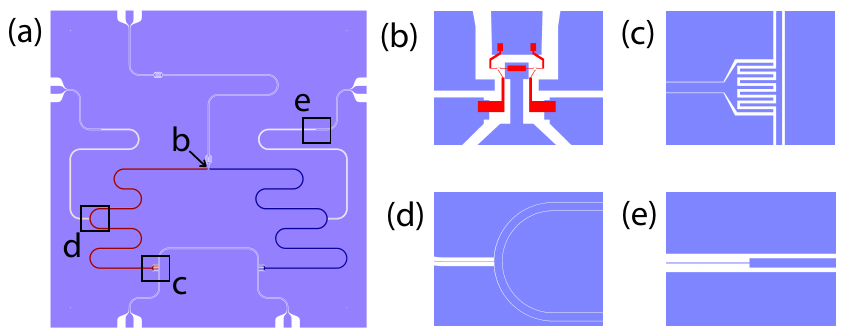}
\caption{The chip-level Nb metalization pattern (a) and close-up views of sites of ground plane discontinuity and changes in line impedance. (b) Common meeting point of the two readout resonators (left and right) and the microwave heating line (from the top). Thin-film shadow mask pattern illustrated in red. (c) Coupling capacitor between feedline and readout resonator. (d) Junction point of the low-$Z$ resonator and high-$Z$ dc injection line. (e) Step-wise transition between the high-$Z$ segment and 50~$\Ohm$ line. Parasitic slot-line modes on the chip are suppressed by about 50 superconducting on-chip bond wires connecting the ground planes across the CPWs, especially at sites (b)-(d) and their mirror positions. \label{Fig:CPW}}
\end{figure}

The coplanar waveguide (CPW) resonators and feedlines are designed using standard practices of microwave engineering~\cite{Simons}. In the case of thick (200 nm) Nb films, the kinetic inductance contribution can be neglected (see below). The width of the central conductor ($S$) and the width of the ground plane gap ($W$) are chosen so as to realize a desired characteristic impedance $(Z_0)$ of the line. One has the relation~\cite{Simons,Barends}
\begin{equation}
Z_0 = \frac{1}{4} \sqrt{\frac{\mu_0}{\epsilon_0 \epsilon_\mathrm{r,eff}}} \frac{K\left(\sqrt{1-k^2}\right)}{K(k)},
\end{equation}
where $\mu_0$ and $\epsilon_0$ are the vacuum permeability and permittivity, respectively, $\epsilon_{\mathrm{r,eff}}$ is the effective dielectric constant, $k = \frac{S}{S+2W}$, and $K$ is the complete elliptic integral of the first kind. For $\epsilon_\mathrm{r,eff}$, we use $(1 + \epsilon_\mathrm{r,Si})/2$, where $\epsilon_\mathrm{r,Si} = 11.68$ is the dielectric constant for Si. The cross-sectional dimensions of the different segments of the resonators and, the predicted characteristic impedances are indicated in Table~\ref{tbl:cpw}. The lengths of the resonator segments are chosen as fractions of the wavelength ($\lambda$), with $\lambda$ calculated as
\begin{equation}
\lambda = \frac{c}{f\sqrt{\epsilon_\mathrm{r,eff}}}.
\end{equation}
The complete network of transmission lines patterned on the device chip is illustrated in Fig.~\ref{Fig:CPW}.

\begin{table}[b]
\begin{tabular}{c|c|c|c}
Segment & $S$ ($\mu$m) & $L$ ($\mu$m) & $Z_0$ $(\Omega)$ \\
\hline
feedline & 20 & 10 & 49.3 \\
resonator, hi-$Z$  & 2 & 19 & 111 \\
resonator, low-$Z$ & 35 & 2.5 & 27.8 \\
\end{tabular}
\caption{Cross-sectional dimensions of the coplanar waveguides used in the sample. \label{tbl:cpw}}
\end{table}

The resonance frequency of CPW resonators is also affected by the kinetic inductance of the superconducting films. To estimate the magnitude of the effect, we use the design formulas from Ref.~\cite{Barends}, and substitute our chosen CPW geometries and typical material parameters for Nb ($\rho = 5~{\mu}\Ohm\,\mathrm{cm}$, $\Delta = 1.5$~meV). The predicted kinetic inductance fraction is approximately 2\,\%. The temperature sensitivity, however, includes a suppression factor $e^{-\Delta/(k_\mathrm{B}T)}$ and is therefore non-existent in the temperature range of our experiment ($T < 300$~mK).

\section{Supplemental data and plots}

\subsection{Comparison of Lorentzian fitting and full theory expressions}

\begin{figure}[t]
\includegraphics[width=86mm]{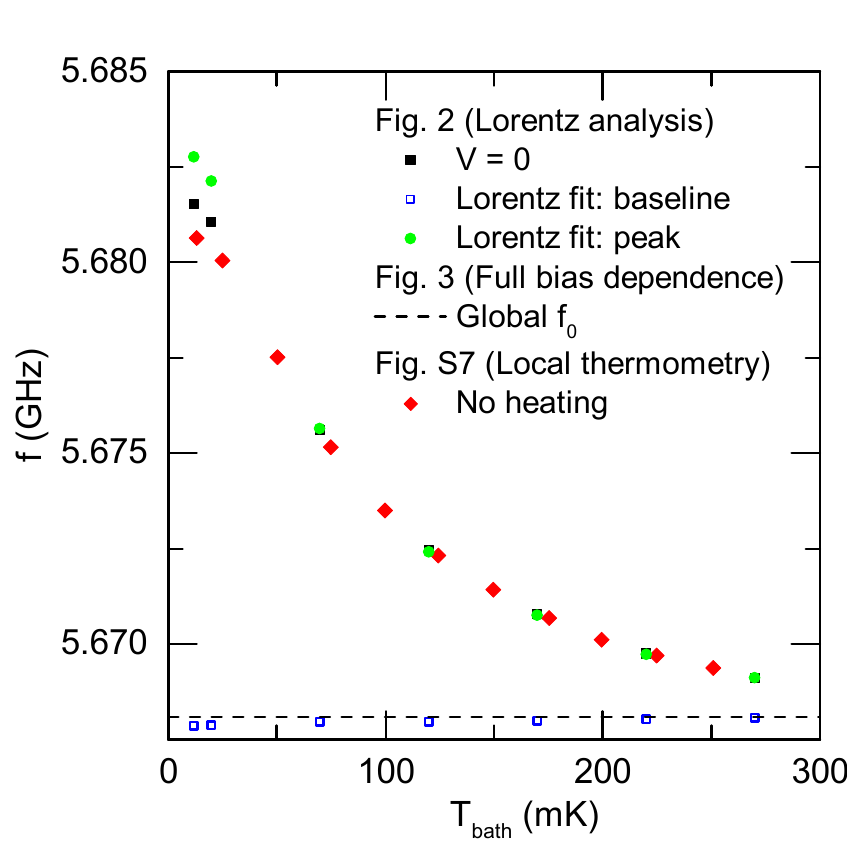}
\caption{Inferred resonance frequencies with different analysis methods. In the data set for local thermometry, the resonance frequency is measured only at $V=0$.\label{Fig:Frequency}}
\end{figure}

The full voltage dependence predicted by the theory is a linear combination of Lorentzian peaks. The behavior close to zero bias is well described by a single Lorentzian function that allows accurate determination of the shunt resistance $R_s$ and the temperature sensed by the junction, as we demonstrate in Fig.~2 of the main text. However, the baseline frequency given by a small voltage range Lorentzian fit (we used $|V| < 2~\mu{V}$) can slightly differ from the true baseline $f_r$ due to presence of wide features that weakly affect the behavior at zero bias. We illustrate the magnitude of this effect in Fig.~\ref{Fig:Frequency} by comparing the Lorentzian fit parameters with the results of the full fits presented in Fig.~3 of the main text. We find a temperature-dependent offset that decreases from 0.7 to 0.25~MHz between base temperature and 270 mK. Note, however, that the Lorentzian fits accurately reproduce the maximal frequency at $V = 0$.

\begin{figure}[t]
\includegraphics[width=86mm]{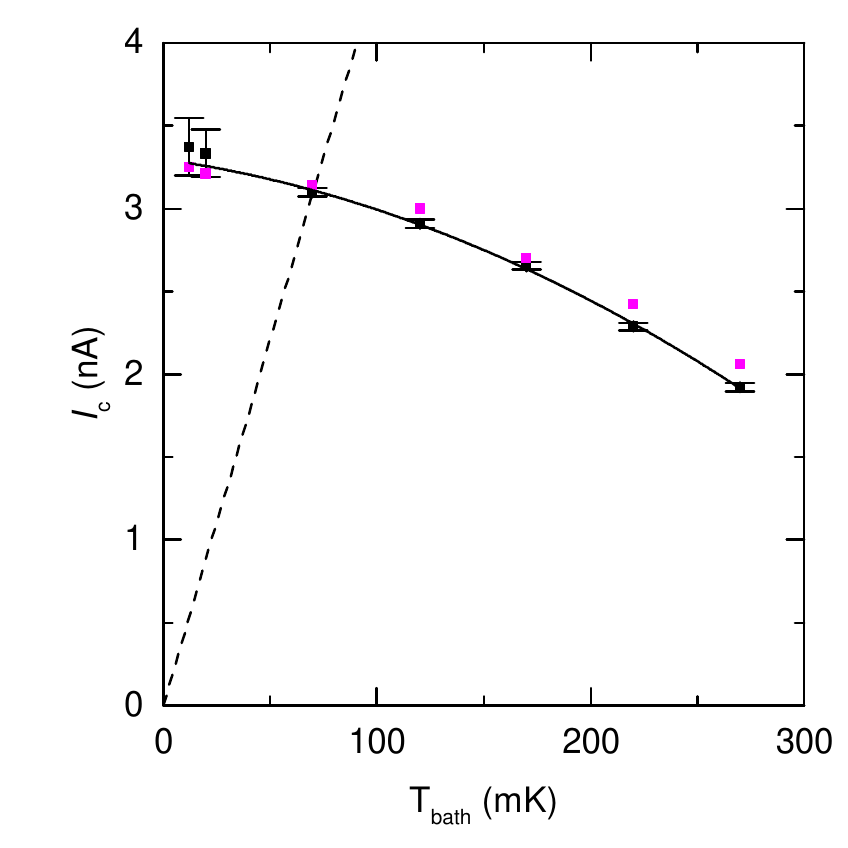}
\caption{Temperature dependence of $I_c$ extracted from a least-squares fit of the Lorentzian zero-bias peak only [black markers, voltage sweeps shown in Fig.~2 (main text)], and from a manual fit considering also the finite-voltage resonances [purple, voltage sweeps shown in Fig.~3 (main text)]. Dashed line is $I_c = 2e k_\mathrm{B} T/\hbar$ and indicates the range of validity of the theory. Solid line is a weighted second order polynomial fit using the estimated uncertainty in $I_c$ as instrumental error.\label{Fig:Ic}}
\end{figure}

Similarly, Eqs.~(10) and (11) of the main text allow determination of the only remaining free parameter in the model, $I_c$, from the product of the Lorentzian width and height parameters ($V_T$ and $\Delta f_T$, respectively). Alternatively, $I_c$ can be fitted by considering the full bias voltage dependence. Due to the uncertainty in the determination of the baseline, the $I_c$ estimates obtained by these two methods can differ by up to 7\%. $I_c$ estimates from both methods indicate that the criterion $I_c \lesssim 2ek_\mathrm{B} T/\hbar$, setting the range of validity for the theory, is met when $T_\mathrm{bath} > 50$~mK (see Fig.~\ref{Fig:Ic}). 

\subsection{Local thermometry}

The fact that the fitted shunt resistance ($R_s = 57.4~\Omega$) is close to the nominal value of the external bias resistor (50~$\Omega$), combined with the very good correspondence between the width of the zero-bias feature and the cryostat temperature, indicates that the external bias resistor dominates the low-frequency electromagnetic environment of the junction. For a simple \emph{non-equilibrium} environment consisting of a series connection of resistors at different temperatures, the effective evironment temperature would be a linear combination of the various temperatures, weighted by the corresponding resistances. Reasoning based on the $R_s$ value alone would yield about 90~\% weight for the external resistor.

Our experiment design allows modulation of the local temperature of the SNS wire(s) through the microwave heating line and by applying a large voltage bias to either of the tunnel junctions. We have performed both of these experiments at various bath temperatures while maintaining the thermometer junction at zero bias and monitoring the resonance frequency. In the case of microwave heating, the wires will be subject to Joule heating given by $\alpha P_\mathrm{mw}$, where $P_\mathrm{mw}$ is the nominal incident microwave power at the sample, and the prefactor $\alpha$ includes power losses due to cabling and imperfect coupling of incident power to the wire. For bias voltages of the order of superconducting gap parameter of the Al leads ($\Delta \approx 200~{\mu}eV$) or larger, the tunnel junctions behave as NIS junctions and the transport is dominated by quasiparticle tunneling. Here, the junction-sourced heat flow to the N electrode can vary from $-0.59 \frac{\Delta^{1/2} (k_\mathrm{B} T)^{3/2}}{e^2 R_\mathrm{T}}$ at $\left|V_\mathrm{NIS}\right| \approx \frac{\Delta - 0.66 k_\mathrm{B} T}{e}$ to Joule heating $V_\mathrm{NIS}^2/(2R_T)$ for $\left|V_\mathrm{NIS}\right| \gg \Delta/e$, where $R_\mathrm{T}$ is the tunnel resistance of the junction~\cite{Nahum,Leivo}. In the present case, the heat flow would be also affected by the modified quasiparticle density of states in the wire, environmental effects described by standard $P(E)$ theory, and possible overheating of the S lead. The steady-state temperature is determined by a thermal balance that includes the electron-phonon heat flow $\Sigma V (T_\mathrm{el}^5 - T_\mathrm{bath}^5$), possibly modified by the density of states affected by proximity effect, and heat flow between the two wires that includes contribution from photonic heat conductance. 

\begin{figure}[t!]
\includegraphics[width=172mm]{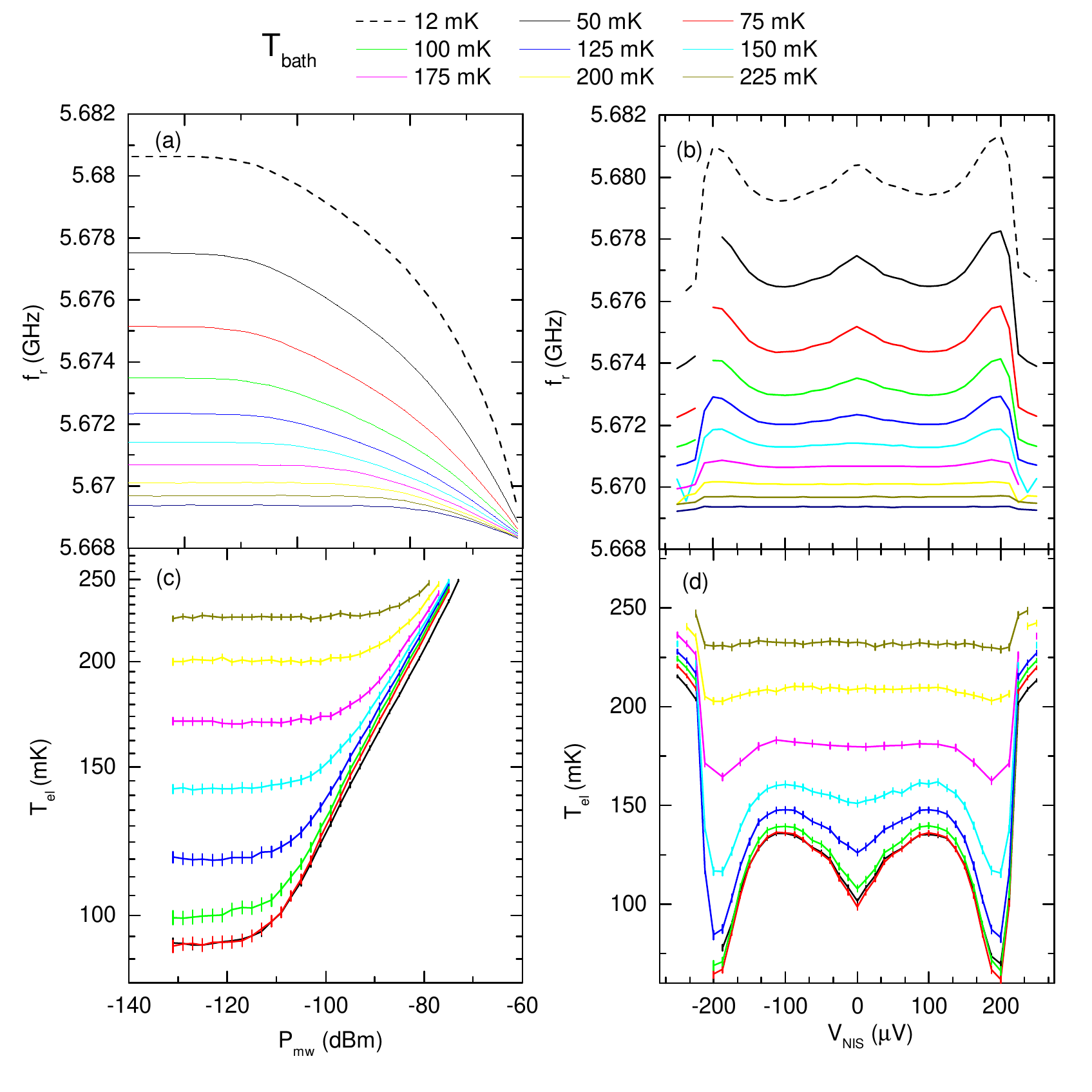}
\caption{Demonstartion of local electron thermometry based on frequency shift. Thermometer resonance frequency at different bath temperatures while the SNS wires are subjected to local heating realized either by microwave radiation from heater line [(a)], or by applying a large bias voltage on the other tunnel junction [(b)]. We decouple the bath and wire temperatures by assuming that the effective temperature of the electromagnetic environment is $T_\mathrm{bath}$, and that the critical current depends on $T_\mathrm{el}$ only. This allows us to estimate the electron temperatures in the range from 75 to 250~mK [(c), (d)]. $T_\mathrm{el}$ estimation was not performed for $T_\mathrm{bath} < 50$~mK, where the theory is expected to fail.\label{Fig:Tel}}
\end{figure}

The observed resonance frequencies under heating [Fig.~\ref{Fig:Tel}(a), (b)] show large variations in response to local temperature modulation with expected shapes (monotonous behavior with respect to $P_\mathrm{mw}$, clear cooling peaks in $V_\mathrm{NIS}$ sweeps). We convert the frequency shifts to $T_\mathrm{el}$ by employing Eqs.~(10) and (11) of the main text with all parameters fixed by the Lorentzian zero-bias fits, and equating the environment temperature to $T_\mathrm{bath}$. $I_c$ is assumed to depend only on $T_\mathrm{el}$ according to the polynomial fit of Fig.~(\ref{Fig:Ic}). Without constructing a detailed thermal model, it is possible to judge the accuracy of the $T_\mathrm{el}$ extraction~[Fig.~\ref{Fig:Tel}(c), (d)] by the magnitude of scatter for large heating powers at different bath temperatures. Strong local heating should drive the wire to a temperature that is largely independent of the surrounding bath temperature. Here, deviations of the order of 30~mK are observed in the upper end of the temperature calibration range, indicating that the modeled separation of $T_\mathrm{bath}$ and $T_\mathrm{el}$ contributions was not entirely accurate. Finally, we note that using a linear combination $(1-a) T_\mathrm{bath} + a T_\mathrm{el}$ with $a > 0$ as the effective environment temperature  does not reduce the scatter.

